\documentclass[prd,superscriptaddress,amsfonts,amssymb,amsmath,showpacs,onecolumn]{revtex4-2}
\usepackage{bm}
\usepackage{amsfonts}
\usepackage{latexsym}
\usepackage[latin1]{inputenc}
\usepackage{graphicx}
\usepackage{amsmath}
\usepackage{palatino}
\usepackage{mathpazo}
\usepackage{textcomp}
\usepackage{mathtools}
\usepackage{float}
\usepackage{booktabs}
\usepackage{dcolumn}
\usepackage{hyperref}
\hypersetup{colorlinks,citecolor=blue}
\usepackage{amsmath}
\usepackage{xcolor}
\usepackage{orcidlink}
\usepackage{caption}
\usepackage{subcaption}
\usepackage{commath}
\newtheorem{theorem}{Theorem}[section]

\newtheorem{lemma}{Lemma}[section]
\newtheorem{corollary}{Corollary}[section]

\newcommand{\be}{\begin{equation}}
\newcommand{\ee}{\end{equation}}
\newcommand{\bea}{\begin{eqnarray}}
\newcommand{\eea}{\end{eqnarray}}
\newcommand{\eeas}{\end{eqnarray*}}
\newcommand{\beas}{\begin{eqnarray*}}

\begin{document}

\title{Impact of curvature based geometric constraints on $F(R)$ theory}

\author{Tee-How Loo\orcidlink{0000-0003-4099-9843}}
\email{looth@um.edu.my}
\affiliation{Institute of Mathematical Sciences, Universiti Malaya, 50603 Kuala Lumpur, Malaysia}
\author{Avik De*\orcidlink{0000-0001-6475-3085}}
\email{avikde@utar.edu.my}
\affiliation{Department of Mathematical and Actuarial Sciences, Universiti Tunku Abdul Rahman, Jalan Sungai Long, 43000 Cheras, Malaysia}
\author{Simran Arora\orcidlink{0000-0003-0326-8945}}
\email{dawrasimran27@gmail.com}
\affiliation{Department of Mathematics, Birla Institute of Technology and
Science-Pilani,\\ Hyderabad Campus, Hyderabad-500078, India.}
\author{P.K. Sahoo\orcidlink{0000-0003-2130-8832}}
\email{pksahoo@hyderabad.bits-pilani.ac.in}
\affiliation{Department of Mathematics, Birla Institute of Technology and
Science-Pilani,\\ Hyderabad Campus, Hyderabad-500078, India.}

\begin{abstract}
Theories of gravity are fundamentally a relation between matter and the geometric structure of the underlying spacetime. So once we put some additional restrictions on the spacetime geometry, the theory of gravity is bound to get the impact, irrespective of whether it is general relativity or the modified theories of gravity. In the present article, we consider two curvature-based constraints, namely the almost pseudo-Ricci symmetric and weakly Ricci symmetric condition. As a novel result, such spacetimes with non-null associated vectors are entirely classified, and then applying the obtained results, we investigate these spacetimes as solutions of the $F(R)$-gravity theory. The modified Friedmann equations are derived and analysed in a model-independent way first. Finally, two $F(R)$ gravity models are examined for recent observational constrained values of the deceleration, jerk, and Hubble parameters. We further discuss the behavior of  energy conditions.
\end{abstract}

\maketitle


\section{Introduction}
The standard gravitational theory led by Einstein's field equations (EFE) $R_{ij}-\frac{R}{2}g_{ij}=\kappa^2 T_{ij},$ showed some limitation in reasonably explaining the cosmic acceleration. To address this issue, researchers tried to modify the Einstein-Hilbert action term containing the Ricci scalar $R$ by an arbitrary function $F(R)$ \cite{Capo/2005} 
\[S=\frac{1}{2\kappa^2}\int F(R) \sqrt{-g}d^4x +\int L_m\sqrt{-g}d^4x,\]
where $L_m$ is the matter Lagrangian producing the stress energy tensor of matter 
\[T_{ij}=-\frac{2}{\sqrt{-g}}\frac{\delta(\sqrt{-g}L_m)}{\delta g^{ij}}.\] 
The $F(R)$-gravity field equations obtained by varying the action $S$ of the gravitational field with respect to the metric tensor $g^{ij}$ read as, 
\be \label{FR}	
F_R(R)R_{ij}-\frac{1}{2}F(R)g_{ij}+(g_{ij}\Box-\nabla_i\nabla_j)F_R(R)=\kappa^2T_{ij},
\ee
where $F_R=\frac{\partial F(R)}{\partial R}$ and $\Box$ represents the d'Alembertian operator. Moreover, we can retrieve the EFEs by setting $F(R)=R$. Here, we assume a perfect fluid type $T_{ij}$ with isotropic pressure $p$ and energy density $\rho$, given by 
\begin{align}\label{eqn:T}
T_{ij}=pg_{ij}+(p+\rho)u_iu_j.
\end{align}
We assume the barotropic equation of state (EoS) $p=\omega \rho$, and the four velocity vector field $u^i$ of the fluid. 

By all means, theory of $F(R)$ gravity is in a mature stage now. Several distinct $F(R)$ forms were  introduced and examined in a variety of contexts. For detailed survey, see \cite{sotiriou/2010a,Girones/2010,Odintsov/2019,Oikonomou/2018} and the references therein. It can very well describe the late time cosmic acceleration, see for details \cite{Barrow/2006,Capozziello/2006,Amendola/2007}. 
In addition to being sufficiently general to encompass some of the fundamental properties of higher-order gravity, $F(R)$ theories of gravity are also unique among higher-order gravity theories in that they appear to be the only ones that can avoid the well-known and catastrophic Ostrogradski instability \cite{Sotiriou/2009}. It is known that $F(R)$ gravity has a particle mode dubbed a 'scalaron', which is explicitly present when $F(R)$ gravity is rewritten in the form of a scalar-tensor \cite{ Deruelle/2008,Nojiri/2006}. In the case of $F(R)$ gravity, one obtains the heavy scalar particles in addition to the graviton when one quantize the scalar field's fluctuations in the background metric. Since the scalar particles in $F(R)$-gravity are large, the pressure might be insignificant, and the strength of the interaction between such scalar particles and the ordinary matter should be feeble, on the order of the gravitational interaction. As a result, such a scalar particle might be an obvious candidate for dark matter. Moreover, $F(R)$ theories have no ghosts and hence can be chosen so that the additional degrees of freedom relative to those of GR do not obviately result in major viability problems. It has also been seen that the certain models exhibit chameleon behavior allowing the theory to have cosmological effects which account for the present acceleration of the universe \cite{Starobinsky/2007}. Furthermore, analysis into the post-Newtonian limit of $F(R)$ demonstrates that the models are in accordance with the solar system tests \cite{Cembranos/2006,Sotiriou/2009}.\\
Despite the fact that the $F(R)$ gravity models provide a possible explanations for the cosmic speed up in the absence of dark energy in the cosmological context, the flexibility in constructing different models of $F(R)$ raises the question of constraining these many possible forms from theoretical and observational perspectives. Energy conditions (ECs) can further put a bound on the model parameter \cite{Santos/2010,Atazadeh/2009,frec/2006,nojiri/2018}.\\
ECs are used to illustrate the spacetime geodesic, attractive behavior, and casual structure. Furthermore, physically, ECs are important techniques for studying black holes and wormholes in different modified gravity \cite{Bamba/2017,Halilsoy/2017, Yousaf/2017}. ECs such as: the strong energy condition (SEC), null energy condition (NEC),  weak energy condition (WEC), and dominant energy condition (DEC) have been used from a theoretical standpoint \cite{Santos/2007,Gong/2007}.  For instance, the Hawking-Penrose singularity theorem uses the WEC and SEC, while the NEC is required to prove the second law of black hole thermodynamics \cite{Carroll/2004}. 

On the other hand, in the differential geometry literature, several geometrical models involving conditions on the Ricci curvature tensor and its covariant or Lie derivative were investigated thoroughly in (pseudo-)Riemannian manifolds. Their interaction with the standard theory of gravity are quite well-versed. Recently, there is a surge in studying such geometric restrictions in the realm of modified gravity theories \cite{aprs, epjp, proj, rip, ps, ucdfr, ucdfr1}. And there is no scope of surprise element in it. Any gravity theory is basically a well-documented revelation of the connection between the matter and the geometry of the underlying spacetime, how it is curved or regarding its other two fundamental geometric entities, namely, the torsion (in the metric teleparallel theory) and the non-metricity (in the symmetric teleparallel theory). So once we put some additional condition on the spacetime geometry, the impact on the field equation is almost ascertained. Two of the most popular and successful spacetime structures among these are almost pseudo Ricci symmetric $(APRS)_n$ and weakly Ricci symmetric $(WRS)_n$, introduced by Chaki and Kawaguchi \cite{aaa} and Tam$\acute a$ssy and Binh \cite{tamsbin93}, respectively.

A non-flat $n$-dimensional pseudo-Riemannian manifold is called an almost pseudo Ricci symmetric spacetime if its Ricci tensor $R_{ij}$
is not identically zero and satisfies the condition 
\be \label{aprs}
\nabla_iR_{jk}=(E_i+A_i)R_{jk}+A_jR_{ki}+A_kR_{ij}, 
\ee
where $A_i$ and $E_i$ are the associated 1-forms. In \cite{ucd1}, the authors studied several examples of an $(APRS)_n$ with non-zero and non-constant Ricci scalar whose conformal curvature tensor vanishes. Under this vanishing conformal curvature condition, $(APRS)_4$ reduces to a Robertson-Walker spacetime \cite{avik}. Similar studies without a cosmological constant were done in \cite{ozen}. In \cite{universe}, the authors considered an almost pseudo Ricci symmetric type FRW universe with a dynamic cosmological constant and equation of state (EoS). $(APRS)_4$ spacetimes were also studied in modified gravity theories. A Robertson-Walker spacetime is $(APRS)_4$ type under certain conditions. Energy conditions were analysed for some popular models of $F(R)$-gravity in $(APRS)_4$ spacetime \cite{aprs}. 

More generally, a weakly Ricci symmetric spacetime is a pseudo-Riemannian manifold whose Ricci tensor satisfies
\be \label{wrs}
\nabla_iR_{jk}=D_iR_{jk}+B_jR_{ki}+A_kR_{ij}. 
\ee
The authors (\cite{deghosh}, \cite{desahanous}) studied the curvature properties in a $(WRS)_n$ with the assumption $B_i\neq A_i$ and with a non-singular Ricci curvature assumption, it was proved that $B_i=A_i$ \cite{mantica}. Recently a $(WRS)_4$ spacetime under general relativistic condition \cite{avik,avikmajhi} and in modified $F(R)$-gravity theory \cite{epjp} for constant Ricci scalar were studied and the energy conditions were explored. 

However, so far, a complete classification of these spacetimes has not been obtained. This motivated us to look into these spacetimes with non-null associated vectors and a flat conformal curvature tensor, as well as their applications in modified $F(R)$ theories of gravity. Similar classification results were obtained recently in two other geometrical models, namely, the pseudo-symmetric spacetimes \cite{rip} and the generalized Ricci-recurrent spacetimes \cite{ps,ijgmmp}.

The following is how the current article is structured: Section \ref{sec2} investigates a 4-dimensional $(APRS)_{4}$ spacetime. We briefly discuss a 4-dimensional $(WRS)_{4}$ spacetime in section \ref{sec3}. In section \ref{sec4}, we investigate a conformally flat $(APRS)_{4}$ spacetime in $F(R)$ theory. In section \ref{sec5}, we observe the behavior of the equation of state parameter. Following this, various ECs are investigated in section \ref{sec6}. Section \ref{sec7} contains the discussions and conclusions.

\section{($APRS)_4$ spacetime} \label{sec2}
In this section we consider an $(APRS)_4$ spacetime, hence from (\ref{aprs}) it follows that 
\be\label{eqn:5}
\nabla_iR_{jk}-\nabla_kR_{ij}=E_iR_{jk}-E_kR_{ij}.
\ee
Contraction over $j$ and $k$ in (\ref{aprs}) gives
\be\label{eqn:2}
\nabla_iR=RE_i+RA_i+2A_lR^l_i.
\ee
Using the identity $\nabla_jR=2\nabla_lR^l_j$, after contraction over $i$ and $k$ in (\ref{aprs}) gives
\be\label{eqn:2b}
\frac12\nabla_jR=E_lR^l_j+RA_j+2A_lR^l_j.
\ee
(\ref{eqn:2})-(\ref{eqn:2b}) give
\be\label{eqn:4}
\nabla_iR=2RE_i-2R_{il}E^l.
\ee
Hence we obtain
\begin{lemma}\label{lem:R=const}
For an $(APRS)_4$ spacetime with nowhere null vector $E^i$, 
the Ricci scalar is a constant if and only if $R_{il}E^l=RE_i$.
\end{lemma}
Suppose that $E_i$ is nowhere null and write    
\be\label{eqn:E}
E_i=\varepsilon\mu u_i; \quad u^lu_l=\varepsilon=\pm 1; \quad \mu\neq0.
\ee
We define the orthogonal operator 
\[
h_{jk}=g_{jk}-\varepsilon u_ju_k; \quad h^{jk}=g^{jk}-\varepsilon u^ju^k.
\]
Then clearly
\begin{align} \label{eqn:h} 	
h^a_jh^{jb}=&h^{ia}h_{ij}h^{jb}=h^{ab}; \quad 
h^l_j=h_{jk}h^{kl}=\delta^l_j-\varepsilon u_ju^l.
\end{align}
We further suppose that the Ricci tensor takes the form
\begin{align}\label{eqn:50}
R_{jk}
=&Pg_{jk}+\varepsilon(R-4P)u_ju_k.      
\end{align}
It follows from (\ref{eqn:4}) and (\ref{eqn:50}) that 
\be\label{eqn:12}
\nabla_iR=\varepsilon \dot Ru_i, \quad \dot R=6\mu P.
\ee
On the other hand, differentiating covariantly (\ref{eqn:50}) gives
\begin{align}\label{eqn:110}
\nabla_iR_{jk}=\nabla_iPh_{jk}+\varepsilon(R-4P)\{u_j\nabla_iu_k+u_k\nabla_iu_j\}
					+\varepsilon(\nabla_iR-3\nabla_iP)u_ju_k.
\end{align}
Comparing (\ref{aprs}) and (\ref{eqn:110}) we have
\begin{align}\label{eqn:112}
\nabla_iPh_{jk}+\varepsilon(R-&4P)\{u_j\nabla_iu_k+u_k\nabla_iu_j\}+\varepsilon(\nabla_iR-3\nabla_iP)u_ju_k \notag\\
=&P\{\varepsilon \mu u_ih_{jk}+A_ih_{jk}+A_jh_{ik}+A_kh_{ij}\} \notag \\
			&+\varepsilon (R-3P)\{\varepsilon \mu u_iu_ju_k+A_iu_ju_k+A_ju_iu_k+A_ku_iu_j\}.
\end{align}
Transvecting with $h^{kl}$, $h^{ia}$ and $h^{jb}$ we have
\begin{align}\label{eqn:122}
h^{ia}\nabla_iPh^{lb}   
=&P\{h^{ia}A_ih^{lb}+h^{jb}A_jh^{la}+h^{kl}A_kh^{ab}\} \notag\\
=&P\{h^{ia}A_ih^{lb}+h^{ib}A_ih^{la}+h^{kl}A_kh^{ab}\} 
\end{align}
where we have used identities in (\ref{eqn:h}).
By switching $a$ and $b$ in (\ref{eqn:122}), we have 
\begin{align}\label{eqn:124}
h^{ib}\nabla_iPh^{la}   
=&P\{h^{ib}A_ih^{la}+h^{ia}A_ih^{lb}+h^{kl}A_kh^{ba}\}. 
\end{align}
Hence 
\[
h^{ia}\nabla_iPh^{lb}   =h^{ib}\nabla_iPh^{la}.   
\]
Contracting over $b$ and $l$ gives
\[
2h^{ia}\nabla_iP=0,   
\]
or 
\be \label{eqn:130}
\nabla_iP=\varepsilon \dot P u_i; \quad (\dot P=u^l\nabla_lP).
\ee
Hence (\ref{eqn:122}) becomes 
\begin{align*}
P\{h^{ia}A_ih^{lb}+h^{ib}A_ih^{la}+h^{ab}A_kh^{kl}\} =0.
\end{align*}
Contracting over $a$ and $b$ gives
\be\label{eqn:140}
5Ph^{il}A_i=0.
\ee
We consider two cases: $P\neq0$ and $P=0$.

\emph{Case (a) $P\neq0$:}
It follows that from (\ref{eqn:140}) that
$$
h^{il}A_i=0.
$$
Since $A_i\neq0$, the previous equation gives 
\be
A_i=\varepsilon  \lambda u_i, \quad (\lambda=A^lu_l\neq 0).
\label{eqn:160}\ee
By applying (\ref{eqn:130})--(\ref{eqn:160}) into (\ref{eqn:112}), we obtain 
\begin{align}\label{eqn:112b}
&\varepsilon \dot Pu_ih_{jk} +\varepsilon (R-4P)\{u_j\nabla_iu_k+u_k\nabla_iu_j\}
				+\varepsilon(\nabla_iR-3\nabla_iP)u_ju_k \notag\\
&=\varepsilon P\{\mu u_ih_{jk}+\lambda u_ih_{jk}+\lambda u_jh_{ik}+\lambda u_kh_{ij}\}
			+\varepsilon (R-3P)(\varepsilon\mu+3\varepsilon\lambda)u_iu_ju_k.
\end{align}
Transvecting with $h^{kl}$ we have
\begin{align}\label{eqn:120}
\dot Pu_ih^l_j+(R-4P)u_j\nabla_iu^l    
=P\{\mu u_ih^l_j+\lambda  u_ih^l_j+\lambda u_jh^l_i\}.
\end{align}
By comparing  the $h^l_j$-components and using (\ref{eqn:130}), we have
\be\label{eqn:22}
\dot P=P(\mu +\lambda); \quad 
\nabla_iP=\varepsilon \dot P u_i. 
\ee
Hence (\ref{eqn:120}) becomes
\[
(R-4P)\nabla_iu^l   =\lambda Ph^l_i.
\]
If $R=4P$, the spacetime becomes an Einstein space; implying $R$ is a nonzero constant.
But we can deduce from Lemma~\ref{lem:R=const} and (\ref{eqn:50}) that $P=R$; contradicting the fact that $R$ is nonzero.
Hence we must have $R\neq 4P$.
It follows that 
\begin{align}\label{eqn:23}
\nabla_iu_k   =\frac{\lambda P}{R-4P}h_{ik}, \quad (\lambda=A^lu_l).
\end{align}
Finally, by applying    (\ref{eqn:22})--(\ref{eqn:23}) into (\ref{eqn:112b}), we obtain 
\begin{align}\label{eqn:180}
\dot R-3\dot P=(R-3P)(\mu +3\lambda); \quad \nabla_i(R-3P)=\varepsilon (\dot R-3\dot P)u_i. 
\end{align}
It follows from  (\ref{eqn:12}), (\ref{eqn:22})--(\ref{eqn:180}) that 
\be\label{eqn:25}
(\dot R-\mu R)\mu=(3\mu R-\dot R)\lambda.
\ee
Next, it follows from (\ref{eqn:22})--(\ref{eqn:23}) and (\ref{eqn:12}) that 
\begin{align*}
\frac{\nabla_j\nabla_iR}{6\varepsilon}
=\nabla_j(\mu Pu_i)			
=&Pu_i\nabla_j\mu +\mu \dot P u_iu_j+\frac{\mu\lambda P^2}{R-4P}h_{ij},
\end{align*}
which implies that $u_i\nabla_j\mu=u_j\nabla_i\mu$ and so
\be\label{eqn:d-mu}
\nabla_i\mu=\varepsilon\dot\mu u_i.
\ee
On the other hand, it follows from (\ref{eqn:23}) that $\dot u_k=u^i\nabla_iu_k=0$.
Hence we can select a suitable local coordinate systems such that 
$u^i=(\partial_t)^i$. 
By virtue of (\ref{eqn:12}), (\ref{eqn:22}), (\ref{eqn:180})--(\ref{eqn:d-mu}), we know that 
\be\label{eqn:func}
P=P(t), \quad R=R(t), \quad \lambda=\lambda(t), \quad \mu=\mu(t)
\ee
are all functions depending on $t$ only.

Let $\mathcal D_1$ be the distribution spanned by $u_i$ and 
$\mathcal D_2$ be its orthogonal complementary distribution, that is,
$\mathcal D_2=\{X_i : X_lu^l=0\}$.
Then both distributions are integrable according to (\ref{eqn:23}) and (\ref{eqn:func}). Moreover, 
$\mathcal D_1$ is authparallel while 
$\mathcal D_2$ is spherical.
As a result, the spacetime is 
is locally a warped product of $\mathbb R$ and three-dimensional (pseudo)-Riemannian manifold $M^*$
and the line element $ds^2$ is locally expressed in the form 
\cite{reck}
\begin{align}\label{eqn:warped}
ds^2=\varepsilon dt^2+a^2(t)g^*_{\mu\nu}dx^\mu dx^\nu
\end{align}
where $1\leq \mu,\nu\leq 3$,  $g^*_{\mu\nu}$ is the metric tensor $M^*$ corresponding  to coordinates  $x^\mu$
and 
\begin{align}\label{eqn:a0}
a(t)=\exp\left(\int^t_{t_0}\frac{\lambda P}{R-4P}dt\right).
\end{align}
By  applying \cite[Corollary 7.43(3)]{oneill} and (\ref{eqn:50}), we see that 
$M^*$ is an Einstein space and since every three-dimensional Einstein space is of constant curvature 
\cite[Proposition 1.120]{besse},
we conclude that the spacetime is locally a warped product of $\mathbb R$ and three-dimensional (pseudo)-Riemannian space form with constant curvature $k$.
Applying the formula for warped product  (\ref{eqn:warped}) gives
\begin{align}
P=&\frac{-\varepsilon a\ddot a-2\varepsilon \dot a\dot a+2k}{a^2} \label{eqn:P}\\
R=&\frac{-6\varepsilon a\ddot a-6\varepsilon \dot a\dot a+6k}{a^2}. \label{eqn:R}
\end{align}
Applying (\ref{eqn:12}) and (\ref{eqn:25}) into (\ref{eqn:a0}) gives
\begin{align}\label{eqn:a}
a(t)=\exp\left(-\int^t_{t_0}\frac{\mu}2\:\frac{\mu R-\dot R}{3\mu R-\dot R}\:\frac{\dot R}{3\mu R-2\dot R}\:dt\right).
\end{align}

\medskip
\emph{Case (b) $P=0$:} In this case, we have  
\be \label{eqn:P=0} R_{ik}=\varepsilon Ru_iu_k.\ee
By Lemma~\ref{lem:R=const}, we have $\nabla_iR=0$. 
Since an $(APRS)_4$ cannot be Ricci-flat, we have $R\neq0$.
Hence  (\ref{eqn:112}) is reduced to 
\begin{align}\label{eqn:112a}
u_j\nabla_iu_k+u_k\nabla_iu_j&=\varepsilon\mu u_iu_ju_k+A_iu_ju_k+A_ju_iu_k+A_ku_iu_j \notag\\
&=\varepsilon(\mu+3\lambda) u_iu_ju_k.
\end{align}
Transvecting with $u^j$ and $u^k$ gives
$$\mu+3\lambda=0.
$$
Hence (\ref{eqn:112a}) becomes
\[
u_j\nabla_iu_k+u_k\nabla_iu_j=0
\]
and so $\nabla_iu_j=0$. This implies that $R^j_iu_j=0$ and so $R=0$ by (\ref{eqn:P=0}); a contradiction to our hypothesis.
Hence this case is impossible. 

Gathering the conclusion in Case (a), together with (\ref{eqn:12}), (\ref{eqn:22}), (\ref{eqn:23}), (\ref{eqn:25})
and (\ref{eqn:P})--(\ref{eqn:a}), yield the following result.

\begin{theorem}\label{thm:QE-APRS}
Suppose  that the $1$-from $E_i$  of an $(APRS)_4$ 
is nowhere null and satisfying 
\[
E_i=\varepsilon\mu u_i; \quad u^lu_l=\varepsilon=\pm 1; \quad \mu\neq0,
\]
and  
the Ricci tensor  satisfies  (\ref{eqn:50}), that is,  
\begin{align*}
R_{jk}
=&Pg_{jk}+\varepsilon(R-4P)u_ju_k.     
\end{align*}
Then 
 the spacetime is locally a warped product of $\mathbb R$ and three-dimensional (pseudo)-Riemannian space form
of constant curvature $k$
whose warped function $a(t)$ satisfies  
\[
a(t)=\exp\left(-\int^t_{t_0}\frac{\mu}2\:\frac{\mu R-\dot R}{3\mu R-\dot R}\:\frac{\dot R}{3\mu R-2\dot R}\:dt\right)
\]
where 
\begin{align*}
A_lu^l=\lambda=&-\mu\frac{\mu R-\dot R}{3\mu R-\dot R}  \\
\frac{\dot R}{6\mu}=P=&\frac{-\varepsilon a\ddot a-2\varepsilon \dot a\dot a+2k}{a^2}
\text{ with } \dot P=P(\mu+\lambda) \\
R=&\frac{-6\varepsilon a\ddot a-6\varepsilon \dot a\dot a+6k}{a^2}.
\end{align*}
\end{theorem}
In particular, when the vector $u^i$ in Theorem~\ref{thm:CF-APRS} is the four-velocity of the fluid,
the spacetime becomes a RW-spacetime. 
\begin{corollary}\label{cor:QE-APRS}
Suppose that  the $1$-from $E_i$  of an $(APRS)_4$ 
is timelike and satisfying 
\[
E_i=-\mu u_i;  \quad \mu\neq0,
\]
where $u^i$ is the four-velocity of the fluid
and  
the Ricci tensor  is of quasi-Einstein type, that is,  
\begin{align*}
R_{jk}
=&Pg_{jk}-
(R-4P)u_ju_k.     
\end{align*}
Then 
 the spacetime is a RW-spacetime whose scale $a(t)$ satisfies  
\[
a(t)=\exp\left(-\int^t_{t_0}\frac{\mu}2\:\frac{\mu R-\dot R}{3\mu R-\dot R}\:\frac{\dot R}{3\mu R-2\dot R}\:dt\right)
\]
where 
\begin{align*}
A_lu^l=\lambda=&-\mu\frac{\mu R-\dot R}{3\mu R-\dot R}  \\
\frac{\dot R}{6\mu}=P=&\frac{a\ddot a+2 \dot a\dot a+2k}{a^2}
\text{ with } \dot P=P(\mu+\lambda) \\
R=&\frac{6a\ddot a+6\dot a\dot a+6k}{a^2}.
\end{align*}
\end{corollary}

We should mention that in \cite{perfect fluid} the authors derived a condition for a perfect fluid spacetime
to be a generalized RW space-time, which in dimension $4$ automatically reduces to a RW spacetime, the converse was also discussed. In \cite{gray} the latter part was discussed, that is, the condition under which a (generalized) RW spacetime becomes a perfect fluid spacetime. 

Next, we consider a conformally flat $(APRS)_4$ spacetime in which the $1$-form  $E^i$ is given by (\ref{eqn:E}).
Then we have 
\be\label{eqn:6}
\nabla_iR_{jk}-\nabla_kR_{ij}=\frac{\nabla_iRg_{jk}-\nabla_kRg_{ij}}6.
\ee
By using (\ref{eqn:5}) and (\ref{eqn:6}), we obtain
\be\label{eqn:7}
6\varepsilon\mu u_iR_{jk}=6\varepsilon\mu u_kR_{ij}+\nabla_iRg_{jk}-\nabla_kRg_{ij}.
\ee
Transvecting with $u^i$ and using (\ref{eqn:4}) give
\be\label{eqn:8}
6\mu R_{jk}=6\varepsilon\mu Ru_ku_j+\dot R g_{jk}-3u_j\nabla_kR-u_k\nabla_jR
\ee
The skew-symmetric part gives
\begin{align*}
3u_j\nabla_kR+u_k\nabla_jR&=3u_k\nabla_jR+u_j\nabla_kR. 
\end{align*}
Hence 
$$
\nabla_jR=\varepsilon \dot Ru_j.
$$
Substituting into (\ref{eqn:8}) we obtain
\begin{align*}
R_{jk}
=&Pg_{jk}+\varepsilon (R-4P)u_ju_k,   
\end{align*}
where $6\mu P=\dot R$.       
By Theorem~\ref{thm:QE-APRS} and Corollary~\ref{cor:QE-APRS}, we have the following results.
\begin{theorem}\label{thm:CF-APRS}
Suppose that the $1$-from $E_i$  of a conformally flat $(APRS)_4$ 
is nowhere null and satisfying 
\[
E_i=\varepsilon\mu u_i; \quad u^lu_l=\varepsilon=\pm 1; \quad \mu\neq0.
\]
Then 
 the spacetime is locally a warped product of $\mathbb R$ and three-dimensional (pseudo)-Riemannian space form
of constant curvature $k$
whose warped function $a(t)$ satisfies  
\[
a(t)=\exp\left(-\int^t_{t_0}\frac{\mu}2\:\frac{\mu R-\dot R}{3\mu R-\dot R}\:\frac{\dot R}{3\mu R-2\dot R}\:dt\right)
\]
where 
\begin{align*}
A_lu^l=\lambda=&-\mu\frac{\mu R-\dot R}{3\mu R-\dot R}  \\
\frac{\dot R}{6\mu}=P=&\frac{-\varepsilon a\ddot a-2\varepsilon \dot a\dot a+2k}{a^2}
\text{ with } \dot P=P(\mu+\lambda) \\
R=&\frac{-6\varepsilon a\ddot a-6\varepsilon \dot a\dot a+6k}{a^2}.
\end{align*}
\end{theorem}

\begin{corollary}\label{cor:CF-APRS}
Suppose that the $1$-from $E_i$  of a conformally flat $(APRS)_4$ 
is timelike and satisfying 
\[
E_i=-\mu u_i;  \quad \mu\neq0,
\]
where $u^i$ is the four-velocity of the fluid.
Then 
 the spacetime is a RW-spacetime whose scale function $a(t)$ satisfies   
\begin{align}\label{eqn:ca}
a(t)=\exp\left(-\int^t_{t_0}\frac{\mu}2\:\frac{\mu R-\dot R}{3\mu R-\dot R}\:\frac{\dot R}{3\mu R-2\dot R}\:dt\right)
\end{align}
where 
\begin{align}
A_lu^l=\lambda=&-\mu\frac{\mu R-\dot R}{3\mu R-\dot R}  \label{eqn:c25}\\
\frac{\dot R}{6\mu}=P=&\frac{a\ddot a+2\dot a\dot a+2k}{a^2} \text{ with } \dot P=P(\mu+\lambda) \label{eqn:c12} \\
R=&\frac{6a\ddot a+6\dot a\dot a+6k}{a^2}. \notag
\end{align}

\end{corollary}

\section{($WRS)_4$ spacetime} \label{sec3}
Let us consider a $(WRS)_4$ spacetime, that is, the Ricci tensor satisfies (\ref{wrs}). Since $\nabla_iR_{jk}=\nabla_iR_{kj}$, we have   
\[
B_jR_{ki}+A_kR_{ij}=B_kR_{ji}+A_jR_{ik}\]
or 
\be\label{eqn:300}
(B_j-A_j)R_{ki}=(B_k-A_k)R_{ji}
\ee
If $B_j=A_j$, then a $(WRS)_4$ becomes an $(APRS)_4$.
Hence we consider $B_j\neq A_j$.
Suppose that $B_j-A_j$ is nowhere null and write    
\be\label{eqn:B-A}
B_j-A_j=\varepsilon(\beta-\alpha) u_j;  \quad \beta-\alpha\neq0.
\ee
where $u^lu_l=\varepsilon=\pm1$, $\alpha=u^lA_l$ and $\beta=u^lB_l$.
We define the orthogonal operator 
\[
h_{jk}=g_{jk}-\varepsilon u_ju_k; \quad h^{jk}=g^{jk}-\varepsilon u^ju^k.
\]
Hence (\ref{eqn:300}) can be simplified as 
\be\label{eqn:300b}
u_jR_{ki}=u_kR_{ji}.
\ee
Transvecting with $u^j$ gives
\be\label{eqn:300c}
\varepsilon R_{ki}=u_kR_{ji}u^j.
\ee
By the symmetry of the Ricci tensor
$$
u_kR_{ji}u^j=u_iR_{jk}u^j.
$$ 
Transvecting with $u^k$ gives
\be\label{eqn:300d}
R_{ji}u^j=\varepsilon u_iR_{jk}u^ju^k.
\ee
By using (\ref{eqn:300c})--(\ref{eqn:300d}), we obtain
\be\label{eqn:310}
R_{ij}=R_{kl}u^ku^lu_iu_j=\varepsilon Ru_iu_j.
\ee
Contracting over $j$ and $k$ in (\ref{wrs}) gives
\be\label{eqn:312}
\nabla_iR=\{D_i+\varepsilon \beta u_i+\varepsilon\alpha u_i\}R.
\ee
On the other hand, differentiating covariantly (\ref{eqn:310}) gives
\begin{align}\label{eqn:320}
\nabla_iR_{jk}=\varepsilon R\{u_j\nabla_iu_k+u_k\nabla_iu_j\}	+\varepsilon\nabla_iRu_ju_k.
\end{align}
Comparing (\ref{wrs}) and (\ref{eqn:320}) we have
\begin{align}\label{eqn:330}
 R&\{u_j\nabla_iu_k+u_k\nabla_iu_j\}+\nabla_iRu_ju_k 
=			R\{D_iu_ju_k+B_ju_iu_k+A_ku_iu_j\}.
\end{align}
Transvecting (\ref{eqn:330}) with $u^j$ and $u^k$, we have
\begin{align*}
\nabla_iR &=R\{D_i+\varepsilon \beta u_i+\varepsilon \alpha u_i\}, \quad (\alpha  =u^lA_l, \beta=u^lB_l).
\end{align*}
Substituting into (\ref{eqn:330})  gives 
$$
 R\{u_j\nabla_iu_k+u_k\nabla_iu_j+\varepsilon\beta u_iu_ju_k+\varepsilon\alpha u_iu_ju_k\}
=			R\{B_ju_iu_k+A_ku_iu_j\}.
$$ 
Since an $(WRS)_4$ cannot be Ricci-flat, we have $R\neq0$ and so
$$
 u_j\{\nabla_iu_k-u_i(A_k-\varepsilon\alpha u_k)\}+ u_k\{\nabla_iu_j-u_i(B_j-\varepsilon\beta u_j)\}=0.
$$ 
This implies that 
\begin{align}\label{eqn:350}
 \nabla_iu_k=u_i(A_k-\varepsilon\alpha u_k)=u_i(B_k-\varepsilon\beta u_k).
\end{align}
Now we suppose further that the $(WRS)_4$ spacetime is conformally flat.
Then by (\ref{wrs}), (\ref{eqn:6}), (\ref{eqn:B-A}) and (\ref{eqn:310}), we obtain
\be\label{eqn:360}
6\varepsilon R\{(D_i-B_i)u_ju_k-(D_j-B_j)u_iu_k
\}=\nabla_iRg_{jk}-\nabla_jRg_{ik}.
\ee
Firstly, transvecting (\ref{eqn:360}) with $h^{jk}$ and $u^i$ to get $3\dot R=0$; then 
transvecting (\ref{eqn:360}) again with $h^{jk}$, we obtain 
\be\label{eqn:370}
2\nabla_iR=0.
\ee
Hence it follows from (\ref{eqn:312}) and (\ref{eqn:360}) that $D_i=B_i$ and 
\[
B_i+\varepsilon\beta u_i+\varepsilon\alpha u_i=0.
\]
We can deduce from this that 
\[
B_i=\varepsilon \beta u_i; \quad 2\beta+\alpha=0.
\]
Applying this to (\ref{eqn:350}) gives 
\[
\nabla_iu_k=0.
\] 
It follows that $R^k_iu_k=0$ and so  $R=0$ by (\ref{eqn:310}); a contradiction to our hypothesis.

Hence we have obtained the following result.

\begin{theorem}\label{thm:CF-WRS}
There does not exist any conformally flat $(WRS)_4$ such that $B^j-A^j$ is a nonzero nowhere null vector field.
\end{theorem}


\section{Application in $F(R)$-gravity} \label{sec4}
This section describes a conformally flat $(APRS)_4$ spacetime with timelike associated vector,
namely $E_i=u_i$,  as a solution of $F(R)$-gravity theory. We have $\varepsilon=\mu=-1$ 
under this formulation. Using relations in Corollary~\ref{cor:CF-APRS},
we compute
\begin{align*}
\nabla_j F_R(R)=&
										-F_{RR}(R)\dot Ru_j\\
\nabla_i\nabla_jF_R(R)=&F_{RRR}(R)\dot R^2u_iu_j+F_{RR}\ddot R u_iu_j-F_{RR}(R)\frac{\dot a}a h_{ij}	\\
\Box F_R(R)=&-F_{RRR}(R)\dot R^2-F_{RR}\ddot R -3F_{RR}(R)\frac{\dot a}a\\
R_{ij}=&-\frac{\dot R}{6}h_{ij}-\frac{6R+3\dot R}6u_iu_j.
\end{align*}
Substituting these relations into the field equations (\ref{FR}) of $F(R)$-gravity theory gives a perfect fluid form of stress energy tensor
\begin{align}
\left(-\frac12F(R)-\frac{\dot R}6F_{R}(R)-\ddot RF_{RR}(R)
		-2\dot R\frac{\dot a}a F_{RR}(R)-\dot R^2F_{RRR}(R)\right)h_{ij} &\notag
\\
+\left(\frac12F(R)-\frac{6R+3\dot R}6F_{R}(R)+3\dot R\frac{\dot a}a F_{RR}(R)\right)u_iu_j &=\kappa^2 T_{ij}.
\end{align}
This gives
\begin{align*}
\kappa^2 p=&-\frac12F(R)-\frac{\dot R}6F_{R}(R)-\ddot RF_{RR}(R)
							-2\dot R\frac{\dot a}a F_{RR}(R)-\dot R^2F_{RRR}(R)\\
\kappa^2\rho=&\frac12F(R)-\frac{6R+3\dot R}6F_{R}(R)+3\dot R\frac{\dot a}a F_{RR}(R).
\end{align*}
Using (\ref{eqn:ca})--(\ref{eqn:c12}), we compute
\begin{align*}
\frac{\dot a}a=&-\frac12\frac{R+\dot R}{3R+\dot R}\:\frac{\dot R}{3R+2\dot R} \\
\ddot R
=&-6\dot P=-6P(\lambda -1)=\dot R(\lambda -1)=\dot R\frac{-2R}{3R+\dot R}.
\end{align*}
These relations bring us to the following result:
\begin{theorem}\label{ec}
In a conformally flat $(APRS)_4$ spacetime solution of $F(R)$-gravity, the energy density $\rho$, isotropic pressure $p$ and the expansion scalar are given by:
\begin{align*}
\kappa^2 p=&-\frac12F(R)-\frac{\dot R}6F_{R}(R)
							+\frac{\dot R}{3R+\dot R}\frac{6R^2+5R\dot R+\dot R^2}{3R+2\dot R}F_{RR}(R)
							-\dot R^2F_{RRR}(R)\\
\kappa^2\rho=&\frac12F(R)-\frac{6R+3\dot R}6F_{R}(R)
							-\frac32\frac{\dot R^2}{3R+\dot R}\frac{R+\dot R}{3R+2\dot R}F_{RR}(R)\\
\nabla_lu^l=&3\frac{\dot a}a=-\frac32\frac{R+\dot R}{3R+\dot R}\:\frac{\dot R}{3R+2\dot R}.
\end{align*}
\end{theorem}
This result is in line with the findings of \cite{epl}, where in the realm of $F(R)$ gravity field equations, the authors considered a conformally flat RW-spacetime a priory, and by introducing the concept of perfect scalar (a scalar field $S$ that satisfies $\nabla_iS=-\dot{S}u_i$) showed a similar perfect fluid form of the stress-energy tensor. Thus the modification of the Einstein-Hilbert action by the inclusion of the term $F(R)=R+f(R)$, say, contributed to the model-specific additional geometric terms corresponding to $f(R)$ which are yet again in a perfect fluid form themselves, resulting into straightforward expressions of the pressure and energy density. We arrive at the same conclusion from a much weaker geometric restriction on the spacetime, namely just the condition (\ref{aprs}) and vanishing Weyl curvature tensor. In contrast, an assumption of RW structure assures flat Weyl curvature, perfect fluid type Ricci curvature, and a timelike, shear-free, and acceleration-free unit four-velocity vector field. 

\section{Cosmographic parameters} \label{sec5}

As is widely known, finding analytical formulations for scale factor and hence predicting the values of the cosmographic parameters is challenging due to the mathematical problems encountered while solving higher-order equations. The choice of the $F(R)$ function has an important role. So, to assess the model's feasibility, it is plausible to assume a parameterized model and contrast it with the data. Hence, it is efficient to incorporate the following functions \cite{Visser/2004,Dabrowski/2005}
\begin{align*}
H &= \frac{1}{a} \frac{da}{dt}, &  q&= -\frac{1}{a} \frac{d^{2}a}{dt^{2}} H^{-2}, &  j&= -\frac{1}{a} \frac{d^{3}a}{dt^{3}} H^{-3}.\\
\end{align*}
which are referred to as the Hubble, deceleration, and jerk parameters, respectively. 
The current values of these factors can be used to characterize the evolutionary phase of the universe. In other words, $q_{0}<0$ denotes accelerated growth, while $j_{0}$ distinguishes between various accelerating models.\\
Differentiating the scalar curvature $R= -6(\dot{H}+2 H^{2})$ as a function of $t$, we obtain:
\begin{align*}
R_{0} &= -6 H_{0}^{2} (1-q_{0}),\\
\dot{R_{0}} &= -6 H_{0}^{3} (j_{0}-q_{0}-2).
\end{align*}

We will see in the following subsection that the cosmological parameters depend only on two observational values, $q_{0}$ and $j_{0}$. As we have examined the $(APRS)_{4}$ structure of spacetime, we can flexibly reduce $\ddot{R}$ to $\dot{R}$ with ease.

\subsection{Model-I}

Assuming the functional form $F(R)= R+ m log(nR)$ \cite{Girones/2010,Amendola/2007}, where $m$, and $n$ are model parameters. It is clear that $F(R)$ is continuous and differentiable for $nR>0$. The best fit model is acceptable with all the measurements and some tensions within the allowed ranges. So, we choose $n<0$. Further, a particular case $m=0$ reduces to the well-accepted general relativity (GR) model. We use the current values of cosmological parameters $q_{0}=-0.55$, $j_{0}=1$, and $H_{0}= 67.9$ km/s/Mpc \cite{Planck/2018,Capo/2019}. 

The obtained representations for pressure $p$ and the energy density $\rho$ are 
\begin{multline}
p= \frac{1}{6} \left(\frac{H_{0} (j_{0}-q_{0}-2) \left(6 H_{0}^2 (q_{0}-1)+m\right)}{q_{0}-1}-3 m \log \left(6 H_{0}^2 n (q_{0}-1)\right)-18 H_{0}^2 (q_{0}-1) \right. \\
\left. -\frac{m (-j_{0}+q_{0}+2) (H (-j_{0}+q_{0}+2)+2 (q_{0}-1))}{H_{0} (q_{0}-1)^2 (2 H_{0} (-j_{0}+q_{0}+2)+3 (q_{0}-1))}-\frac{2 m (-j_{0}+q_{0}+2)^2}{(q_{0}-1)^3}\right),
\end{multline}

\begin{multline}
\rho= \frac{1}{6} \left(\frac{3 (H_{0} (j_{0}-q_{0}-2)-2 q_{0}+2) \left(6 H_{0}^2 (q_{0}-1)+m\right)}{q_{0}-1} 
 +3 \left(m \log \left(6 H_{0}^2 n (q_{0}-1)\right)+6 H_{0}^2 (q_{0}-1)\right) \right. \\
 \left. +\frac{3 m (-j_{0}+q_{0}+2)^2 (H (-j_{0}+q_{0}+2)+q_{0}-1)}{2 (q_{0}-1)^2 (H_{0} (-j_{0}+q_{0}+2)+3 (q_{0}-1)) (2 H_{0} (-j_{0}+q_{0}+2)+3 (q_{0}-1))}\right).
\end{multline}


\subsection{Model-II}
We use the functional form of $F(R)= \alpha Exp(\beta/R)-R$ \cite{Girones/2010,Amendola/2007}, where $\alpha$, and $\beta$ are model parameters. Here, $\alpha=0$ reciprocates to a well motivated general relativity (GR) case. We use the present values of cosmological parameters as $q_{0}=-0.55$, $j_{0}=1$, and $H_{0}= 67.9$ km/s/Mpc \cite{Planck/2018,Capo/2019}.

The pressure $p$ and the energy density $\rho$ read as
\begin{multline}
p=   \frac{1}{1296}\alpha  e^{\frac{\beta }{6 H_{0}^2 (q_{0}-1)}} \left(\frac{6 \beta  (-j_{0}+q_{0}+2) (H_{0} (-j_{0}+q_{0}+2)+2 (q_{0}-1)) \left(\beta +12 H_{0}^2 (q_{0}-1)\right)}{H_{0}^5 (q_{0}-1)^4 (2 H_{0} (-j_{0}+q_{0}+2)+3 (q_{0}-1))}+ \frac{36 \beta  (-j_{0}+q_{0}+2)}{H_{0} (q_{0}-1)^2} \right.\\
\left. +\frac{\beta  (-j_{0}+q_{0}+2)^2 \left(\beta ^2+216 H_{0}^4 (q_{0}-1)^2+36 \beta  H_{0}^2 (q_{0}-1)\right)}{H_{0}^6 (q_{0}-1)^6}-648\right) +H_{0}^2 (H_{0} (-j_{0}+q_{0}+2)+3 (q_{0}-1)),
\end{multline}

\begin{multline}
\rho= 3 H_{0}^2 (H_{0} (-j_{0}+q_{0}+2)+q_{0}-1)+\frac{1}{144} \alpha  e^{\frac{\beta }{6 H_{0}^2 (q_{0}-1)}} \left(\frac{12 \beta  (H_{0} (-j_{0}+q_{0}+2)+2 (q_{0}-1))}{H_{0}^2 (q_{0}-1)^2}\right. \\
\left. -\frac{\beta  (-j_{0}+q_{0}+2)^2 (H_{0} (-j_{0}+q_{0}+2)+q_{0}-1) \left(\beta +12 H_{0}^2 (q_{0}-1)\right)}{H_{0}^4 (q_{0}-1)^4 (H_{0} (-j_{0}+q_{0}+2)+3 (q_{0}-1)) (2 H_{0} (-j_{0}+q_{0}+2)+3 (q_{0}-1))}+72\right).
\end{multline}

One can determine the constraints on model parameters using the positive behavior of density for both models. Also, it is seen that the pressure behaves negatively, which causes the equation of state $\omega=\frac{p}{\rho}$ to be negative. Henceforth, the negative equation of state depicts the accelerated expansion of the universe.

\section{Energy conditions} \label{sec6}

In this section, we will study various energy conditions (ECs) and their cosmological effects in modified $F(R)$ gravity \cite{Wang/2010}. The energy-momentum tensor $T_{\mu \nu}$ describes the energy-momentum distribution and stress caused by matter or other non-gravitational fields. Different ECs include the null (NEC), weak (WEC), strong (SEC), and dominant energy condition (DEC). These conditions are developed using Raychaudhuri equations describing the timelike, spacelike, or lightlike curves of gravity congruence and attractive nature. Raychaudhuri equations are then read as \cite{Kar/2007,Carroll/2004}
\begin{eqnarray}
\frac{d\theta}{d\tau} &=& -\frac{1}{2} \theta^{2} -\sigma_{\mu \nu}\sigma^{\mu \nu}+ \omega_{\mu \nu}\omega^{\mu \nu} - R_{\mu \nu} k^{\mu}k^{\nu}, \\
\frac{d\theta}{d\tau} &=& -\frac{1}{3} \theta^{2} -\sigma_{\mu \nu}\sigma^{\mu \nu}+ \omega_{\mu \nu}\omega^{\mu \nu} - R_{\mu \nu} u^{\mu}u^{\nu}.
\end{eqnarray}

where  $\omega_{\mu \nu}$, $\sigma^{\mu \nu}$, $\theta$, are the rotation, shear, and expansion, respectively, corresponding to the congruences defined by the null vector $k^{\mu}$ and timelike vector $u^{\mu}$, respectively. The Raychaudhuri equation shows that for any spatial shear tensor with $\sigma^{2}= \sigma_{\mu \nu} \sigma^{\mu \nu} \geq 0$, and for any hypersurface (which is orthogonal congruence) that imposes $\omega_{\mu \nu}=0$, the attractiveness condition, namely $\frac{d\theta}{d\tau}<0$ gives $R_{\mu \nu} k^{\mu}k^{\nu}\geq 0$. The preceding conditions can be rewritten as $T_{\mu \nu} k^{\mu}k^{\nu}\geq 0$ in terms of stress- energy tensor.\\
Accordingly, ECs are categorized as follows: The NEC is a result of  $T_{\mu \nu} k^{\mu}k^{\nu}\geq 0$, which leads to the well-known $\rho+p \geq 0$ form. The SEC is obtained by the positive behavior of the timelike vector, which gives $\rho+3p \geq 0$. The violation of the SEC represents the accelerated expansion of the universe. In addition to NEC, WEC requires the positivity of the energy density for any observer at any point, that is, $\rho \geq 0$, $\rho+p \geq 0$. At last, DEC results in $\rho \geq 0$, $\rho\pm p \geq 0$. \\
One can check that the NEC, DEC validate their conditions using the expressions of pressure and density in section \ref{sec5}. Since the pressure in the above section is negative, so it contributes more to the SEC violation for both models.
In certain cases, a violation of ECs may occur without the system is unacceptable. The effectiveness of the inflationary theory, as well as notable findings of cosmic acceleration, refer to the universe violating SEC, which is demonstrated in the current scenario for both the models.


\section{Final Remarks} \label{sec7}

New gravity theories produced by modifying Einstein's general theory of relativity, provide an alternative explanation for current cosmic acceleration without requiring the existence of an extra spatial dimension or an exotic component of dark energy. $F(R)$ gravity is one such motivated modified gravity theory,
the simplest of the lot. Past few years of cosmological research reveals that it is not difficult to construct $F(R)$ gravity models that lead to a desired cosmological background evolution. In particular, from a given form of scale factor $a(t)$, we can always reconstruct the $F(R)$ model to produce the required evolutionary stage. However, for the purpose of preciseness in the data syncing and to keep the minute details in the cosmological evolution intact, we still have to go beyond the simple $F(R)$ models. A stable de-Sitter spacetime solution, Big-Bang Nucleosynthesis, the growth of cosmological perturbation and structure formation, all these phenomena can rule out most of the models and lead to strong viability constraints \cite{sotiriou/2010a}. The present work, on the other hand discusses some constraints on spacetime geometry to restrict the viable $F(R)$ models. On any spacetime, starting from a simple yet popular differential geometric condition on the covariant derivative of the Ricci curvature tensor, namely the $(APRS)_4$ and a vanishing Weyl curvature tensor, we have reproduced Friedmann type equations in $F(R)$ theory which is valid in any conformally flat $(APRS)_4$ spacetime. The snap parameter may be relevant for studies involving higher redshifts since cosmic data allows us to measure up to the third order derivative of the scale factor with respect
to time. For the time being, no reliable measurements of the snap parameter have been reported. If we look closely, $F(R)$ theory involves snap parameter
due to its third order derivative of Ricci scalar $R$ \cite{Santos/2007}, we have efficiently removed this dependency. It is seen that $(APRS)_4$ spacetime do not involve
the higher order derivatives of $R$. In due process, we have also succeeded to completely classify the geometric structures $(APRS)_4$ and $(WRS)_4$. Also to note that we have not imposed any further additional strict constraints like vanishing or constant Ricci scalar, unlike many of the existing literature, which usually makes things much simpler. 
\\

We examine the accelerated expansion using two functional forms $F(R)= R+ m log(n R)$, where $m$ and $n$ are model parameters and $F(R)= \alpha Exp(\beta/R)-R$, where $\alpha$, and $\beta$ are model parameters. A comprehensive study of these two models in a $(APRS)_{4}$ spacetime is carried out, revealing insight into the characteristics of density, pressure, and the EoS parameter. It is observed that the EoS parameter shows negative behavior depicting acceleration in the universe.  Moreover, in this setting, the present observational values of the Hubble, deceleration and jerk parameters are utilized to study various energy conditions for constraining the model parameters. The accepted inflationary theory, as well as the recent observations of cosmic acceleration refer to a non-validation (violation) of the SEC in the universe. As a result, it can be seen that NEC and DEC validate their developed conditions through positive behavior. However, the SEC can be seen violating its condition suggesting the accelerated expansion of the universe.

\section{Data availability} No datasets were generated or analysed during the current study.

\section{Acknowledgements}

AD and LTH are supported by the FRGS grant (Grant No. FRGS/1/2021/STG06/UTAR/02/1). SA acknowledges CSIR, New Delhi, India for Senior Research Fellowship. PKS acknowledges CSIR, New Delhi, India for financial support to carry out the Research project[No.03(1454)/19/EMR-II Dt.02/08/2019]. We are very much grateful to the honorable referee and to the editor for the illuminating suggestions that have significantly improved our work in terms of research quality, and presentation.


\end{document}